\documentclass[12pt,a4paper]{article}

\usepackage{a4wide}
\usepackage{amsmath}
\usepackage{bm}
\usepackage{amssymb}
\usepackage{graphicx}
\usepackage{epsfig}
\usepackage{cite}
\usepackage{epic}

\newcommand{\als}{{a_s}}

\def\w{\omega}
\def\CAP#1{{{C_A^#1}}}
\def\CFP#1{{{C_F^#1}}}

\def\nfP#1{{{n_f^#1}}}
\def\z#1{{{\zeta_#1}}}
\def\alsP#1{{{a_s^#1}}}
\def\aw#1{{{\bm{[\omega^{-#1}]}}}}

\newcommand{\cNS}{{\mathcal {N}}=4\; \mathrm {SYM}}
\newcommand{\NS}{{\mathrm {NS}}}
\newcommand{\CA}{{C_A}}
\newcommand{\CF}{{C_F}}
\newcommand{\nf}{{n_f}}
\newcommand{\TF}{{T_F}}

\newcommand{\cN}{{\cal N}}

\begin{document}

\begin{flushright}\footnotesize
\texttt{HU-EP-14/62}\\
\vspace{0.5cm}
\end{flushright}

\begin{center}
{\Large{\bf
Generalised double-logarithmic equation in QCD}}
\vspace{15mm}

{\sc
V.~N.~Velizhanin}\\[5mm]

{\it Institut f{\"u}r  Mathematik und Institut f{\"u}r Physik\\
Humboldt-Universit{\"a}t zu Berlin\\
IRIS Adlershof, Zum Gro\ss{}en Windkanal 6\\
12489 Berlin, Germany\\
and
}\\
{\it Theoretical Physics Division\\
Petersburg Nuclear Physics Institute\\
Orlova Roscha, Gatchina\\
188300 St.~Petersburg, Russia}\\[5mm]

\textbf{Abstract}\\[2mm]
\end{center}

\noindent{
We present a generalisation of the double-logarithmic equation for the anomalous dimension of the non-singlet unpolarized twist-2 operators in QCD. Using the known three-loop result, this generalisation allows to predict a small $x$ expansion of the four-loop non-singlet splitting functions in QCD for all powers of logarithms up to the single-logarithm term.
}

\newpage

\setcounter{page}{1}

The double-logarithmic equation was originally formulated during the study of the asymptotic behavior of the scattering amplitudes in Quantum Electrodynamics~\cite{Frolov:1966hn,Gorshkov:1966ht,Gorshkov:1966qd}. With the help of the method, proposed by Sudakov~\cite{Sudakov:1954sw} for the evaluation of Feynman integrals, the authors of Refs.~\cite{Frolov:1966hn,Gorshkov:1966ht,Gorshkov:1966qd} derived a Bethe-Saltpeter equation, which sums the leading logarithm terms (in this case $(\alpha_{\mathrm{em}}\ln^2\!x)$) in all orders of perturbation theory.
Employing the same methods, investigations of other applicability regions  of the resummation procedure were performed in Quantum Chromodynamics (QCD) 
and the famous 
Dokshitzer-Gribov-Lipatov-Altarelli-Parizi (DGLAP)~\cite{Gribov:1972ri,Altarelli:1977zs,Dokshitzer:1977sg} and Balitsky-Fadin-Kuraev-Li\-pa\-tov (BFKL)~\cite{Kuraev:1977fs,Balitsky:1978ic} equations were obtained.

Then, in the papers of Kirschner and Lipatov~\cite{Kirschner:1982qf,Kirschner:1982xw,Kirschner:1983di} a new approach for the resummation of the double-logarithmic terms was proposed.
This new method allows to perform the double-logarithmic resummation for different amplitudes and channels, considerably extending the conventional approach.
The main point of the new approach is based on isolating the softest virtual particle with the lowest transverse momentum in the graphs.
The authors of Ref.~\cite{Kirschner:1983di} proposed a set of equations for the partial waves of the amplitudes for different channels in the double logarithmic approximation.
In general, these equations are ordinary differential equations of Riccati type and in some cases they are just algebraic ones. The study of the double-logarithmic equation~\cite{Kirschner:1983di} provided an information about behaviour of the structure function in the region of small $x$~\cite{Ermolaev:1995fx,Blumlein:1995jp,Blumlein:1997em,Ermolaev:2000sg}. Moreover, it was extended to Standard Model~\cite{Fadin:1999bq} and to (super)gravity theory~\cite{Bartels:2012ra}.

Although the DGLAP and the BFKL equations were studied in the higher-order approximations in QCD~\cite{Floratos:1977au,GonzalezArroyo:1979df,Moch:2004pa,Fadin:1998py}, the double-logarithmic equation has not yielded any general results beyond the leading-logarithm approximation. In this work, we propose such a generalisation, which was originally discovered during the investigation of the analytical properties of the anomalous dimension of the twist-2 operators in $\cN=4$ Supersymmetric Yang-Mills (SYM) theory~\cite{Velizhanin:2011pb}.

The double-logarithmic equation for the non-singlet anomalous dimension of the twist-2 operator in QCD near $j=0+\omega$, which can be obtained from the corresponding equation for the amplitudes given in Ref.~\cite{Kirschner:1983di}, can be written as:
\begin{equation}\label{DLKL}
\gamma_{\NS}(\omega)=-2\,\CF\frac{\als}{\omega}+\frac{1}{\omega}\Big(\gamma_{\NS}(\omega)\Big)^2\,,
\end{equation}
where
\begin{equation}
\gamma(j)=\sum_{\ell}\alsP{\ell}\gamma^{(\ell-1)}(j)\,,\qquad\qquad \als=\frac{\alpha_s}{4\pi}=\frac{g^2}{16\pi^2}
\end{equation}
and the  anomalous dimension $\gamma(j)$ has poles for all non-positive $j=0,-1,-2,\ldots$ being the function of the nested harmonic sums, defined as (see Refs.~\cite{Vermaseren:1998uu,Blumlein:1998if}):
\begin{eqnarray}
S_a (M)\ =\ \sum^{M}_{j=1} \frac{(\mbox{sign}(a))^{j}}{j^{\vert a\vert}}\, , \qquad
S_{a_1,\ldots,a_n}(M)\ =\ \sum^{M}_{j=1} \frac{(\mbox{sign}(a_1))^{j}}{j^{\vert a_1\vert}}
\,S_{a_2,\ldots,a_n}(j)\, .\label{vhs}
\end{eqnarray}
The double-logarithmic equation~(\ref{DLKL}) provides the information about the highest poles in $\omega$ in all orders of perturbation theory through a trivial solution:
\begin{equation}
\gamma(0+\omega)=-\frac{\omega}{2}+\frac{\omega }{2} \sqrt{1-8\,\CF\frac{\als}{\omega^2}}=-2\,\CF\frac{\als}{\omega}-4\,\CFP2\frac{\alsP2}{\omega^3}-16\,\CFP3\frac{\alsP3}{\omega^5}-80\,\CFP4\frac{\alsP4}{\omega^7} 
+\ldots\label{DLKLSolExp}
\end{equation}
The double-logarithmic equation~(\ref{DLKL}) sums all terms, which are proportional to $(\als\ln^2x)^k$ in all orders of perturbation theory in case we transform the result~(\ref{DLKLSolExp}) into $x$-space by applying
\begin{equation}\label{wlnx}
{\mathcal M}\Big[\ln^k x\Big](N)=(-1)^k\frac{k!}{N^{(k+1)}}\ .
\end{equation}
The double-logarithmic equation gives the important information about the leading behavior of the splitting function at small $x$. However, this leading order result requires corrections. Such corrections can be taken into account through calculations of splitting and coefficient functions within perturbation theory, which are known up to next-to-next-to-leading order in QCD so far~\cite{Floratos:1977au,GonzalezArroyo:1979df,Moch:2004pa,Vermaseren:2005qc}. But there is no any other extension of the double-logarithmic equation~(\ref{DLKL}) beyond the leading-logarithm approximation.

Such generalisation was discovered in the maximally extended $\cN=4$ SYM theory~\cite{Velizhanin:2011pb}, where the anomalous dimension for twist-2 operators are know at this moment up to seven loops~\cite{Kotikov:2003fb,Kotikov:2004er,Kotikov:2007cy,Bajnok:2008qj,Lukowski:2009ce,Marboe:2014sya,Marboe:2016igj}.
By performing the analytic continuation for these results, that can be easily done with the help of \texttt{HARMPOL} package~\cite{Remiddi:1999ew} for \texttt{FORM}~\cite{Vermaseren:2000nd}, we can study the changes of the original double-logarithmic equation~(\ref{DLKL}) with the expansion of the anomalous dimension in the orders of perturbation theory $g$ and parameter $\omega$. Such work was started by L.N. Lipatov and A. Onishchenko for even $j=0,-2,-4,-6,\ldots$ in 2006, but was not published, then, some improvement of the double-logarithmic equation was proposed by L.N. Lipatov\cite{Kotikov:2007cy}. Surprisingly, that in the most simple case $j=0+\omega$ the generalisation was found in $\cN=4$ SYM theory in a very simple form~\cite{Velizhanin:2011pb}:
\begin{equation}\label{DLgener}
\gamma_{{}_{\cNS}}(2\,\omega+\gamma_{{}_{\cNS}})=\sum_{k=1}\sum_{m=0}{\mathfrak C}_m^k\,\omega^m\,g^{2k}\,,
\end{equation}
where the right-hand side is regular in $\omega$.
The solution of the generalised double-logarithmic equation~(\ref{DLgener}) gives corrections to the leading-logarithm approximation~(\ref{DLKLSolExp}), that is, if we know anomalous dimension in $\ell$ loops we know the information about all poles up to $(\als/\omega^2)^k\omega^{2\ell}$ in all orders of perturbation theory. As poles in $\omega$ correspond to $\ln x$ through Eq.~(\ref{wlnx}) we know the resummation of the logarithmic terms in all orders of perturbation theory up to $(\als^k\ln^{2k-2(\ell-1)}\!x)$ term or in the ${\mathrm {N}}^{2(\ell-1)}{\mathrm{LLA}}$ approximation.

For QCD we know the full non-singlet anomalous dimension up to three loops at this moment~\cite{Gross:1973ju,Georgi:1951sr,Floratos:1977au,GonzalezArroyo:1979df,Moch:2004pa}. So, we can expand the result for the anomalous dimension near $j=0+\omega$, which looks like (we suppress the colour factor $\TF  = 1/2$ in $SU(N_c)$)
\begin{eqnarray}
\frac{\gamma^{(0)}(\omega)}{\CF}&=&
-\frac{2}{\w}
-1
+ 2(2 {\z2}-1)\,\w
+ 2(1-2 {\z3})\,\w^2
+ 2(2 {\z4}-1)\,\w^3
+ 2(1-2 {\z5})\,\w^4\,,\label{ADNSQCDDLExpL1}
\end{eqnarray}
\begin{eqnarray}
\frac{\gamma^{(1)}(\omega)}{\CF}&=&
-\frac{4 }{\w^3}  \CF
+\frac{1}{\w^2}\bigg[
\frac{22}{3} \CA 
-\frac{4}{3} \nf
-4 \CF
\bigg]
+\frac{1}{\w}\bigg[
\Big(8 {\z2}+4\Big) \CF
-\frac{302}{9} \CA
+\frac{44}{9} \nf
\bigg]\nonumber\\&&
+\bigg(12 {\z3}+\frac{421}{18}\bigg) \CA
-\frac{29}{9} \nf
+\bigg(-16 {\z3}-\frac{19}{2}\bigg) \CF\nonumber\\&&
+\w \bigg[
\CA \bigg(\frac{268 {\z2}}{9}-\frac{124  {\z3}}{3}-{\z4}-\frac{170}{9}\bigg)
+\nf \bigg(-\frac{40 {\z2}}{9}+\frac{16 {\z3}}{3}+\frac{20}{9}\bigg)\nonumber\\&&
+\CF \Big(16-8 {\z2}+32 {\z3}-14 {\z4}\Big)
\bigg]
+\w^2 \bigg[
\CA \bigg(\frac{104}{9}-\frac{160 {\z3}}{9}+65 {\z4}-40 {\z5}\bigg)\nonumber\\&&
+\nf \bigg(\frac{40 {\z3}}{9}-8 {\z4}-\frac{8}{9}\bigg)
+\CF \Big(-32 {\z2} {\z3}+16 {\z2}-54 {\z4}+104 {\z5}-28\Big)\bigg]\,,\label{ADNSQCDDLExpL2}
\\
\frac{\gamma^{(2)}(\omega)}{\CF}&=&
-\frac{16}{\w^5} \CFP2
+\frac{1}{\w^4}\bigg[
\CF \bigg(44 \CA-8 \nf\bigg)
-24 \CFP2
\bigg]\nonumber\\[.3mm]&&
+\frac{1}{\w^3}\bigg[
(208 {\z2}-8) \CFP2
+\bigg(-192 {\z2}
-\frac{944}{9}\bigg)\CF\CA
+\frac{128}{9} \CF\nf
+\frac{88}{9} \CA \nf\nonumber\\[.3mm]&&
+\bigg(60 {\z2}
-\frac{242}{9}\bigg) \CAP2
-\frac{8}{9} \nf^2
\bigg]
+\frac{1}{\w^2}\bigg[
(-192 {\z2}
-96 {\z3}
-30) \CFP2\nonumber\\[.3mm]&&
+\bigg(216 {\z2}
+48 {\z3}
+\frac{370}{9}\bigg) \CF\CA
-\frac{88 }{9}\CF\nf
+\bigg(8 {\z2}
-\frac{1268}{27}\bigg)\CA \nf\nonumber\\[.3mm]&&
+\bigg(-92 {\z2}
+\frac{3934}{27}\bigg) \CAP2
+\frac{88}{27} \nf^2
\bigg]
+\frac{1}{\w}\bigg[
(308 {\z2}+192 {\z3}-324 {\z4}+62) \CFP2\nonumber\\[.3mm]&&
+\bigg(268 {\z4}
-\frac{532}{9}
-{\z2} \frac{1304}{9}
-{\z3} \frac{808 }{3}
\bigg)  \CA\CF
+\frac{325}{9} \nf\CF
\nonumber\\[.3mm]&&
+\bigg(112 {\z2}
+72 {\z3}
-81 {\z4}
-\frac{9737} {27}\bigg) \CAP2
-\frac{32 }{9}\nfP2
\nonumber\\[.3mm]&&
+\bigg(-\z2\frac{160}{9}
-\z3\frac{32 }{3}\bigg) \nf\CF
+\bigg(-8 {\z2} +16 {\z3} +\frac{2474}{27}\bigg)\CA \nf
\bigg]\nonumber\\[.3mm]&&
+\bigg(448 {\z3} {\z2}-560 {\z2}-316 {\z3}+96 {\z4}+304 {\z5}-\frac{57}{2}\bigg) \CFP2\nonumber\\[.3mm]&&
+\bigg(
-480 {\z2} {\z3}
-160{\z5}
-\frac{1951}{36}
+{\z2} \frac{1672}{3}
+{\z4} \frac{86}{3}
-{\z3}\frac{332}{9}
\bigg) \CF \CA
\nonumber\\[.3mm]&&
+\bigg(
-\frac{2}{9}
-{\z2} \frac{64}{3}
+{\z4} \frac{28}{3}
+{\z3} \frac{344}{9}
\bigg)\nf
\nonumber\\[.3mm]&&
+\bigg(16 {\z2} -\frac{80 {\z3}}{3}-\frac{14 {\z4}}{3}-\frac{670}{9}\bigg)\CA \nf
+\frac{83}{27} \nf^2\nonumber\\[.3mm]&&
+\bigg(
\frac{31087}{108}
-176 {\z2}
+144 {\z2} {\z3}
+\frac{500 {\z3}}{3}
-\frac{133 {\z4}}{3}
-60 {\z5}
\bigg) \CAP2\label{ADNSQCDDLExpL3}
\end{eqnarray}
and substitute the obtained expressions into the original double-logarithmic equation~(\ref{DLKL}).
We have found, that it does change in a minimal way if we add also the QCD $\beta$-function in the left-hand side\footnote{$\beta$-function in $\cN=4$ SYM theory is equal to zero in all orders of perturbation theory.}\footnote{There is a difference between the normalization of the anomalous dimension in $\cN=4$ SYM theory and in~QCD, which produces a difference in the left-hand sides of eqs.~(\ref{GDLEqQCDL3}) and~(\ref{DLgener}).}:
\begin{eqnarray}
\gamma_{{\NS}}(\omega+\gamma_{{\NS}}-\beta/\als)
&=&
\als \bigg\{
-2
-\w
+ (4 {\z2}-2)\w^2
+ (2-4 {\z3})\w^3
+ (4 {\z4}-2) \w^4
\bigg\}\CF\nonumber\\[1.3mm]&&\hspace*{-24.4mm}
+\alsP2 \bigg\{
\w^2 \bigg[
\CA \CF \bigg(\frac{268 {\z2}}{9}-56 {\z3}-{\z4}-\frac{104}{9}\bigg)
+\CF \nf \bigg(\frac{8}{9}-\frac{40 {\z2}}{9}+8 {\z3}\bigg)\nonumber\\[1.3mm]&&\hspace*{-18.4mm}
+\CFP2 (-24 {\z2}+40 {\z3}+10 {\z4}+24)\bigg]
-\frac{335}{9} \CA \CF
+\frac{50}{9} \CF \nf
+(13-8 {\z2}) \CFP2\nonumber\\[1.3mm]&&\hspace*{-18.4mm}
+\w \bigg[
\bigg(\frac{44 {\z2}}{3}+12 {\z3}+\frac{289}{18}\bigg)\CA \CF
-\bigg(\frac{8 {\z2}}{3}+\frac{17}{9}\bigg) \CF \nf
-\bigg(8 {\z2}+\frac{27}{2}\bigg) \CFP2
\bigg]
\bigg\}\nonumber\\[1.3mm]&&\hspace*{-24.4mm}
+\alsP3 \bigg\{
\frac{1}{\w^2}\bigg[144 {\z2} \CFP3-192 {\z2} \CA \CFP2+60 {\z2} \CAP2 \CF\bigg]\nonumber\\[1.3mm]&&\hspace*{-18.4mm}
+\frac{1}{\w}\bigg[
8 {\z2} \CA \CF \nf
+304 {\z2} \CA \CFP2
-92 {\z2} \CAP2 \CF
-16 {\z2} \CFP2 \nf
-240 {\z2} \CFP3\bigg]\nonumber\\[1.3mm]&&\hspace*{-18.4mm}
+\CA \CF \nf \bigg(\frac{608}{9}-8 {\z2}+8 {\z3}\bigg)
+\CAP2 \CF \bigg(112 {\z2}+116 {\z3}-81 {\z4}-\frac{15455}{54}\bigg)\nonumber\\[1.3mm]&&\hspace*{-18.4mm}
+\CFP2 \nf \bigg(\frac{352{\z2}}{9}-\frac{32 {\z3}}{3}+\frac{154}{9}\bigg)
+\CFP3 (340 {\z2}+128 {\z3}-140 {\z4}+1)\nonumber\\[1.3mm]&&\hspace*{-18.4mm}
-\frac{38}{27} \CF \nfP2
+\CA \CFP2 \bigg(\frac{1771}{18}-\frac{4792 {\z2}}{9}-\frac{736{\z3}}{3}+272 {\z4}\bigg)
\bigg\}\label{GDLEqQCDL3}\\[2mm]&&\hspace*{-29mm}
=4\alsP3   {\z2}\CF  \Big(\CA-2 \CF\Big)\bigg[
\frac{3\big(5 \CA-6 \CF\big)}{\w^2}
-\frac{\big(23 \CA-30\CF-2\nf\big)}{\w}\bigg]+{\mathcal O}(\omega^0)\,,\qquad\label{GDLEqQCDL3SP}
\end{eqnarray}
where the coefficients for the $\beta$-function in QCD
\begin{equation}
\label{rengroup}
\beta(a_s)  =  -\beta_0 a_s^2 - \beta_1 a_s^3-\beta_2 a_s^4
\end{equation}
are the following up to the three-loop order~\cite{Gross:1973id,Jones:1974mm,Tarasov:1980au}:
\begin{eqnarray}
\label{eq:beta3}
\beta_{0} & = &  \frac{11}{3} \CA - \frac{2}{3} \nf\,,
\\
 \beta_{1} & = &
 \frac{34}{3}\CAP2 - 2 \CF \nf -\frac{10}{3} \CA \nf\,,
\\
 \beta_{2} & = &  \frac{2857}{54} \CAP3
 +\Big(\CFP2 - \frac{205}{18} \CF \CA - \frac{1415}{54} \CAP2\Big) \nf
 + \Big(\frac{11}{9} \CF
  + \frac{79}{54} \CA\Big) \nfP2\,. \label{mainbeta}
\end{eqnarray}

One can see that the terms on the right-hand side of Eq.~(\ref{GDLEqQCDL3SP}), which has poles in $\omega$, are proportional to $\z2$ and $(\CA-2\CF)$ colour structure. Therefore they are suppressed by the subcolour factor $(\CA-2\CF)=1/N_c$ for $SU(N_c)$.
If we assume, that the modification of the original double-logarithmic equation~(\ref{DLKL}) will contain pole terms, which are proportional only to $\z2$ or $(\CA-2\CF)$, we will obtain the resummation of the logarithm in all orders of perturbation theory in the form of the following solution:
\begin{equation}
\gamma_{\NS}^{{\mathrm {N}}^{2(\ell-1)}{\mathrm{LLA}}}(\omega)=-\frac{\omega-\beta/\als}{2}
+\frac{\omega-\beta/\als}{2} \sqrt{1+\frac{4}{\big(\omega-\beta/\als\big)^2}\sum_{\ell=1}\sum_{m=0}{\mathfrak D}_m^\ell\,\omega^m\,\alsP{\ell}}\ ,\label{GDLQCDSolution}
\end{equation}
where coefficients ${\mathfrak D}_m^k$  can be read directly from Eq.~(\ref{GDLEqQCDL3}) and we should drop out all the terms proportional to $\z2$ or/and suppressed by the colour factor $(\CA-2\CF)$ in ${\mathfrak D}_m^3$.

\begin{figure}[th]
\includegraphics[scale=1]{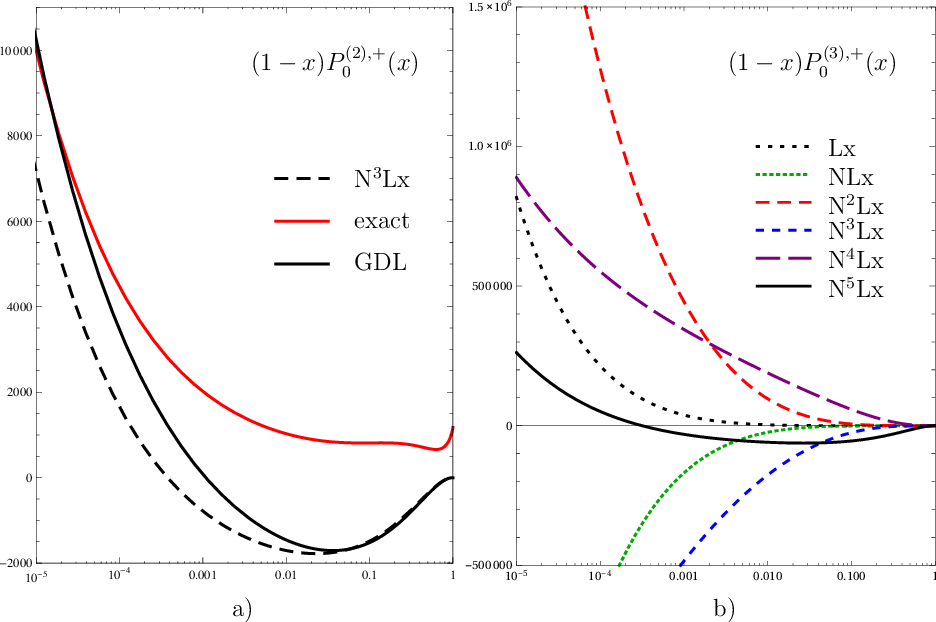}
\caption{
The $\nf$-independent contributions of $P^{(2),+}_{0}(x)$ and  $P^{(3),+}_{0}(x)$ to the splitting function $P^{+}_{\mathrm{ns}}(x)$,
multiplied by $(1-x)$. In the left part the exact result\cite{Moch:2004pa} is compared to the small-$x$ approximation and our solution~(\ref{GDLQCDSolution}). In the right part the pre\-dic\-tions for the small-$x$ approximations to four-loop $P^{(3),+}_{0}(x)$ from our solution~(\ref{GDLQCDSolution}) are presented.
\label{fig1}}
\end{figure}

Let's transform the solution~(\ref{GDLQCDSolution}) into the logarithmic form in terms of $\ln x$ with the help of Eq.~(\ref{wlnx}) and study its properties.
We start with the comparison of our result Eq.~(\ref{GDLQCDSolution}) with the exact result from the three-loop calculations~\cite{Moch:2004pa}.
Expanding the solution~(\ref{GDLQCDSolution}) only with ${\mathfrak D}_m^1$ and ${\mathfrak D}_m^2$ (or using only two-loop results~(\ref{ADNSQCDDLExpL1}) and~(\ref{ADNSQCDDLExpL2}))
we found, that the expansion of $\hat P^{(2)}_{+,0}(x)$ near $x=0$, which can be written in general case as
\begin{equation}
\hat P^{(2),+}_{x\to 0}\!(x)=\hat D_0^{(2),+}\ln^4\! x+\hat D_1^{(2),+}\ln^3\! x+\hat D_2^{(2),+}\ln^2\! x+\hat D_3^{(2),+}\ln x\,,
\end{equation}
differs from the full result in Eq.~(4.15) Ref.~\cite{Moch:2004pa} in the following terms
\begin{eqnarray}
D_2^{+}-\hat D_2^{(2),+} &=& -6\z2(\CA-2\CF)(5\CA-6\CF)\CF\,, \\[1mm]
D_1^{+}-\hat D_1^{(2),+} &=& -4\z2(\CA-2\CF)(23\CA-30\CF-2\nf)\CF \,.
\end{eqnarray}
We compare the obtained result with Fig.(2) from Ref.~\cite{Moch:2004pa}, employing the same input data.
It is clear, that the difference between $\mathrm{N}^3\mathrm{Lx} $ approximation and the exact result in Fig.(\ref{fig1}a) is a constant for the small values of $x$.
Our result (the black solid line in Fig.(\ref{fig1}a)), obtained from the two loops (or using only ${\mathfrak D}_m^1$ and ${\mathfrak D}_m^2$), is different up to this constant and the term, which is proportional to $\z2(\CA-2\CF)$. One can see, that we have a very good agreement for small $x$ with the exact result.

Expanding the solution~(\ref{GDLQCDSolution}) with ${\mathfrak D}_m^1$, ${\mathfrak D}_m^2$ and ${\mathfrak D}_m^3$ (or using three-loop results~(\ref{ADNSQCDDLExpL1})-(\ref{ADNSQCDDLExpL3}))
we found, that the expansion of $\hat P^{(3),+}_{x\to 0}(x)$ near $x=0$, which has the following general form
\begin{equation}
\hat P^{(3),+}_{x\to 0}\!(x)=\hat D_0^{(3),+}\ln^6\! x+\hat D_1^{(3),+}\ln^5\! x+\hat D_2^{(3),+}\ln^4\! x+\hat D_3^{(3),+}\ln^3\! x+\hat D_4^{(3),+}\ln^2\! x+\hat D_5^{(3), +}\ln x
\end{equation}
can be written as:
\begin{eqnarray}
\hat D_0^{(3),+} &=&
\frac{\CFP4}{9}\,,\\
\hat D_1^{(3),+}& =&
\frac{22}{9} \CA \CFP3-\frac{4}{9} \CFP3 \nf-\frac{4}{3} \CFP4\,,\\
\hat D_2^{(3),+} &=&
\Big[\frac{16}{3}-\frac{56 {\z2}}{3}\Big]\CFP4 -\frac{44}{9} \CA \CFP2 \nf+\frac{170}{9} \CA \CFP3+\frac{121}{9} \CAP2 \CFP2-\frac{20}{9} \CFP3 \nf+\frac{4}{9} \CFP2 \nfP2 \nonumber\\&&
-2 {\aw2} {\z2} \CF^2 \Big(5 \CA-6 \CF \Big) \Big(\CA-2 \CF \Big)\,,
\end{eqnarray}
\begin{eqnarray}
\hat D_3^{(3),+} &=&
\frac{44}{27} \CA \CF \nfP2-\frac{2092}{27} \CA \CFP2 \nf-\frac{242}{27} \CAP2 \CF \nf+\CA \CFP3 \Big[\frac{10}{3}-176 {\z2}+48 {\z3}\Big]\nonumber\\&&
+\frac{6530}{27} \CAP2 \CFP2+\frac{1331}{81} \CAP3 \CF+\Big[32 {\z2}-\frac{16}{3}\Big] \CFP3 \nf+\frac{152}{27} \CFP2 \nfP2-\frac{8}{81} \CF \nfP3\nonumber\\&&
+\CFP4 \Big[96 {\z2}-\frac{320 {\z3}}{3}-\frac{106}{3}\Big]
-\frac{8}{3} {\aw1} {\z2} \CF^2 \Big(\CA-2 \CF \Big) \Big(23 \CA-30 \CF-2 \nf \Big)
\nonumber\\&&
-\frac{2}{3} {\aw2} {\z2} \CF \Big(5 \CA-6 \CF \Big) \Big(\CA-2 \CF \Big) \Big(11 \CA-6 \CF-2 \nf \Big)\,,
\\
\hat D_4^{(3),+} &=&
\CA \CFP2 \nf \Big[\frac{1024 {\z2}}{9}-64 {\z3}-\frac{32968}{81}\Big]+\CA \CFP3 \Big[98-\frac{1024 {\z2}}{3}+\frac{416 {\z3}}{3}-532 {\z4}\Big]\nonumber\\&&
+\frac{644}{27} \CA \CF \nfP2-\frac{1390}{9} \CAP2 \CF \nf+\CAP2 \CFP2 \Big[\frac{114740}{81}-\frac{4436 {\z2}}{9}+32 {\z3}+162 {\z4}\Big]\nonumber\\&&
+\frac{25003}{81} \CAP3 \CF+\CFP3 \nf \Big[\frac{352 {\z2}}{3}+\frac{256 {\z3}}{3}-\frac{170}{3}\Big]+\Big[\frac{2144}{81}-\frac{80 {\z2}}{9}\Big] \CFP2 \nfP2 \nonumber\\&&
+\CFP4 \Big[-760 {\z2}-256 {\z3}+1360 {\z4}-112\Big]
-\frac{88}{81} \CF \nfP3
\nonumber\\&&
-\frac{2}{3} {\aw1} {\z2}\CF \Big(\CA-2 \CF \Big) \Big(23 \CA-30 \CF-2 \nf \Big) \Big(11 \CA-6 \CF-2 \nf \Big)
\nonumber\\&&
+24 {\aw2} {\z2} (2 {\z2}-1) \CF^2 \Big(5 \CA-6 \CF \Big) \Big(\CA-2 \CF \Big)\,,\\
\hat D_5^{(3),+} &=&
\CFP4 \Big[3072 \z2\z3-1232 \z2-944 \z3-1576 \z4+1920 \z5-130\Big]
\nonumber\\&&
+\CA \CFP3 \Big[\frac{332\z2}{3}-2496 \z2 \z3-\frac{12448\z3}{3}+3964\z4-960\z5-\frac{2761}{3}\Big]
\nonumber\\&&
+\CAP2 \CFP2 \Big[576 \z2\z3-\frac{14776\z2}{9}+\frac{8984\z3}{3}-\frac{3922\z4}{3}-240\z5+\frac{254225}{81}\Big]
\nonumber\\&&
+\CAP3 \CF \Big[\frac{146482}{81}-\frac{1232\z2}{3}-264\z3+297\z4\Big]
\nonumber\\&&
+\CFP3 \nf \Big[\frac{712\z2}{3}+\frac{2080\z3}{3}-536\z4+\frac{500}{3}\Big]
+\CA \CF \nfP2 \Big[\frac{7561}{81}-\frac{16 {\z2}}{3}+\frac{32 {\z3}}{3}\Big]
\nonumber\\&&
+\CA \CFP2 \nf \Big[\frac{5072\z2}{9}-\frac{1328\z3}{3}+\frac{472\z4}{3}-\frac{90538}{81}\Big]
-\frac{64}{27} \CF \nfP3
\nonumber\\&&
+\CAP2 \CF \nf \Big[104 \z2-\frac{32\z3}{3}-54\z4-\frac{64481}{81}\Big]
+\CFP2 \nfP2 \Big[\frac{7736}{81}-\frac{448{\z2}}{9}\Big]
\nonumber\\&&
+16 {\aw1} {\z2} (2 {\z2}-1) \CF^2 \Big(\CA-2 \CF \Big) \Big(23 \CA-30 \CF-2 \nf \Big)
\nonumber\\&&
+48 {\aw2} {\z2} (2 {\z3}-1) \CF^2 \Big(5 \CA-6 \CF \Big) \Big(\CA-2 \CF \Big) \,,
\end{eqnarray}
where all terms, that are proportional to $\bm{[\omega^{-2}]}$ and  $\bm{[\omega^{-1}]}$ come from the first and the second terms in Eq.~(\ref{GDLEqQCDL3SP}) correspondingly.

Inserting $\CA = 3$ and $\CF = 4/3$ and numerical values of $\z2,\ \z3$ and $\z5$ one can find
\begin{eqnarray}
\hat D_0^{(3),+} &=& 0.351166\,,\\
\hat D_1^{(3),+} &=& 13.1687-1.0535 n_f\,,\\
\hat D_2^{(3),+} &=&
269.244 -31.3416 n_f+0.790123 n_f^2-13.6469 \aw2\,,\\
\hat D_3^{(3),+} &=&
   2818.6
   -408.66 n_f
   +16.5267 n_f^2
   -0.131687 n_f^3
   +\aw2 \left(6.8234 n_f-85.293\right)\nonumber\\&&\hspace*{40mm}
   +\aw1 \left(5.1988 n_f-75.383\right)\,,
   \\
\hat D_4^{(3),+} &=&
   17395.
   -2869.9 n_f
   +116.470 n_f^2
   -1.44856 n_f^3
   +375. \aw2\nonumber\\&&\hspace*{40mm}
   + \aw1 \left(-1.94955 n_f^2+52.638 n_f-353.36\right)\,,
\\
\hat D_5^{(3),+} &=&
   54450.4
   -9381.3 n_f
   +413.8 n_f^2
   -3.16049 n_f^3
   +459.9 \aw2\nonumber\\&&\hspace*{40mm}
   + \aw1 \left(1035.7-71.4 n_f\right).
\end{eqnarray}
This means that the contributions from uncontrolled terms $\aw2$ and $\aw1$ are small.
These results are shown in Fig.(\ref{fig1}b).

In conclusion we would like to note, that the generalised double-logarithmic equation~(\ref{DLgener}), obtained in $\cN=4$ SYM theory~\cite{Velizhanin:2011pb}, provides us with a new information about resummation in QCD. The generalised double-logarithmic equation for QCD~(\ref{GDLEqQCDL3}) is violated only by terms, which are proportional to $\z2(\CA-2\CF)$ in Eq.~(\ref{GDLEqQCDL3SP}). We hope, that one can find the origin of these terms to restore Eq.~(\ref{DLgener}). Note, also, that recently the result for the four-loop  non-singlet anomalous dimension  was obtained in the planar limit~\cite{Moch:2017uml}, when $\CA=N_c$ and $\CF=N_c/2$ and their combination $(\CA-2\CF)=0$. The cancellation of this term restores our generalised double-logarithmic equation for QCD~(\ref{GDLEqQCDL3SP}) and it was confirmed in Ref.~\cite{Moch:2017uml}, that it is correct in the four-loop orders too.

\bigskip

\noindent
{\large\bf {Acknowledgments}}

I would like to thank L.N. Lipatov and A. Onishchenko for useful discussions.
This research is supported by a Marie Curie International Incoming Fellowship within the 7th European Community Framework Programme, grant number PIIF-GA-2012-331484, by DFG SFB 647 ``Raum -- Zeit -- Materie. Analytische und Geometrische Strukturen'' and by RFBR grant 13-02-01246-a.

\end{document}